\documentclass[aip,graphicx]{revtex4-1}
\usepackage{siunitx}
\usepackage{graphicx}
\usepackage{amsmath}
\usepackage{float}
\usepackage{xcolor}

\draft 

\begin{document}


\title{Temperature Measurements of Liquid Flat Jets in Vacuum} 

\newcommand*\geaffilcmd{GAP-Biophotonics, Universit{\'e} de Gen{\'e}ve, 1205 Geneva, Switzerland}
\newcommand*\ethaffilcmd{Laboratorium für Physikalische Chemie, ETH Z{\"u}rich, 8093, Z{\"u}rich, Switzerland}

\author{Yi-Ping Chang}
\affiliation{\geaffilcmd}

\author{Zhong Yin}
\email{yinz@ethz.ch}
\affiliation{\ethaffilcmd}

\author{Tadas Balciunas}
\affiliation{\geaffilcmd}
\affiliation{\ethaffilcmd}

\author{Hans Jakob W{\"o}rner}
\affiliation{\ethaffilcmd}

\author{Jean-Pierre Wolf}
\affiliation{\geaffilcmd}


\begin{abstract}
Sub-\SI{}{\micro\metre} thin samples are essential for spectroscopic purposes. The development of flat micro-jets enabled novel spectroscopic and scattering methods for investigating molecular systems in the liquid phase. However characterization of the temperature of these ultra-thin liquid sheets in vacuum has not been systematically investigated. Here we present a comprehensive temperature characterization of two methods producing sub-micron flatjets, using optical Raman spectroscopy: colliding of two cylindrical jets and a cylindrical jet compressed by a high pressure gas. Our results reveal the dependence of the cooling rate on the material properties and the source characteristics, i.e. nozzle orifice size, flowrate, pressure. We show that materials with higher vapour pressures exhibit faster cooling rates which is illustrated by comparing the temperature profile of liquid water and ethanol flatjets. In a sub-\SI{}{\micro\metre} liquid sheet, the temperature of the water sample reaches around 268 K and the ethanol around 253 K.
\end{abstract}


\maketitle 

\section{Introduction}

X-ray absorption spectroscopy (XAS) is a powerful method for the investigation of fundamental electronic properties of matter. However, the use of XAS in the soft X-rays is experimentally challenging in bulk materials, as transmission limits the sample thickness to some hundreds of nanometers.\cite{smith2017soft} This is especially true in liquids, for which geometrical effects and differences in absorption lengths can, in addition, lead to artifacts in the measured absorbance spectra.\cite{eisebitt1993determination, busse2020} Ultrathin liquid samples are also very valuable for other spectroscopic techniques like ultrafast MeV electron diffraction, which has the advantage of shorter wavelength and stronger interaction with matter than X-rays.\cite{nunes2020liquid, Yang2020} For MeV electron diffraction experiments in liquid samples, ultra-thin homogeneous samples are required to minimize noise contribution from inelastic electron scattering events. The small penetration depths of even high energetic electrons are $<$1 \SI{}{\micro\metre} and thin samples with around 800 nm thickness are necessary to avoid multiple scattering events.\cite{deponte2011towards, akar2006electron} 

Several methods have been developed to generate a sub-micron flat sheet in a vacuum environment, namely, two colliding cylindrical jets, 3D printed nozzles and gas-dynamic jets.\cite{Ekimova2015, Galinis2017, Koralek2018} Earlier works of colliding jets have been performed under ambient conditions showing a thickness down to a \SI{}{\micro\metre}.\cite{Taylor1960,miller1960distribution, huang1970break, bremond2006atomization} Recent reported thicknesses reach 100 nm for both colliding jets and gas-dynamic jet.\cite{Ekimova2015,Galinis2017,Koralek2018, yin2020} These submicron-thickness flatjets have enabled several novel achievements in science, such as investigating liquid samples using high-order harmonic spectroscopy\cite{Luu2018, Svoboda21}, fs-transient XAS\cite{Smith2020}, angular photoelectron spectroscopy\cite{malerz2021setup}, and fs-electron scattering in liquids\cite{lin2021imaging}. Despite these advantages and applications, an essential property of flatjets in vacuum, i.e. its temperature, remains mostly unknown.

Liquid samples in vacuum experience fast evaporation that result in a temperature decrease potentially reaching down to the supercooled regime.\cite{sellberg2014ultrafast,goy2018shrinking, wilson2004investigation, charvat2004} For a majority of molecular systems, the temperature can have a vital impact on the properties and the evolution of the system. Although temperature characterization of super-cooled water droplets and cylindrical jets have been conducted in previous studies \cite{Faubel1988,Smith2006,goy2018shrinking,Grisenti2018} using Raman spectroscopy, systematic investigation of liquid flatjets has not been performed yet. In this work, we present measurements of the temperature profiles of liquid water (H$_2$O) and ethanol (C$_2$H$_5$OH) in flatjets in vacuum using Raman spectroscopy. A systematic investigation of the temperature dependence on the material vapor pressure, nozzle orifice size, flowrate, and initial temperature of the liquid sample is presented.

\section{Experimental methods}

We investigated two types of flatjets: (1) a collision-based (impingement) flatjet formed by the collision of two cylindrical liquid jets\cite{Ekimova2015}, and (2) a jet produced by a microfluidic nozzle, in which an initially cylindrical liquid jet is strongly compressed sideways by helium gas.\cite{Koralek2018} These two configurations are the two most widely used setups for liquid-phase XAS measurements. The determination of the temperature profiles within these liquid jets is therefore of key importance for spectroscopic investigations in the X-ray domain.

In both setups, we use a high-pressure liquid-chromatography (HPLC) pump to deliver the samples to the nozzles.\cite{Smith2020} The vacuum in the experimental chamber typically ranges between $5\times 10^{-3}$ to $1\times 10^{-2}$ millibar during the measurements. In order to keep acceptable vacuum conditions, the sample is captured in a cold trap at liquid-nitrogen temperature. The investigated samples are distilled water and absolute pure ethanol with at least 99.8\,\% purity (Merck). 

For the collision-based flatjet\cite{Ekimova2015,Luu2018,Smith2020}, two cylindrical liquid jets from two quartz nozzles of equal inner diameters (either \SI{18}{\micro\metre} or \SI{60}{\micro\metre}) collide at an angle of 48$^{\circ}$ and generate a flatjet with a thickness that can be as thin as $500$\,\SI{}{\nano\metre}, as assessed by white-light interferometry. 

The thickness depends on several parameters, such as the orifice size of the nozzle, flowrate, type of liquid, and the measurement position. For the presented measurements, the flowrate was 2 ml/min for \SI{18}{\micro\metre} nozzles and 5 ml/min for \SI{60}{\micro\metre} nozzles, with a maximum jet velocity of 65.5m/s and 14.7 m/s respectively.

For the gas-compressed flatjet, a microfluidic gas-dynamic chip nozzle from Micronit Microtechnologies BV with an orifice of \SI{50}{\micro\metre} was used.\cite{Koralek2018} The focusing gas was helium and its pressure was adjusted using a regulator between 5 and 10 bars. The liquid sample flowrate was 1 ml/min.

\begin{figure}[htbp] 
\centering
\includegraphics[width=10cm]{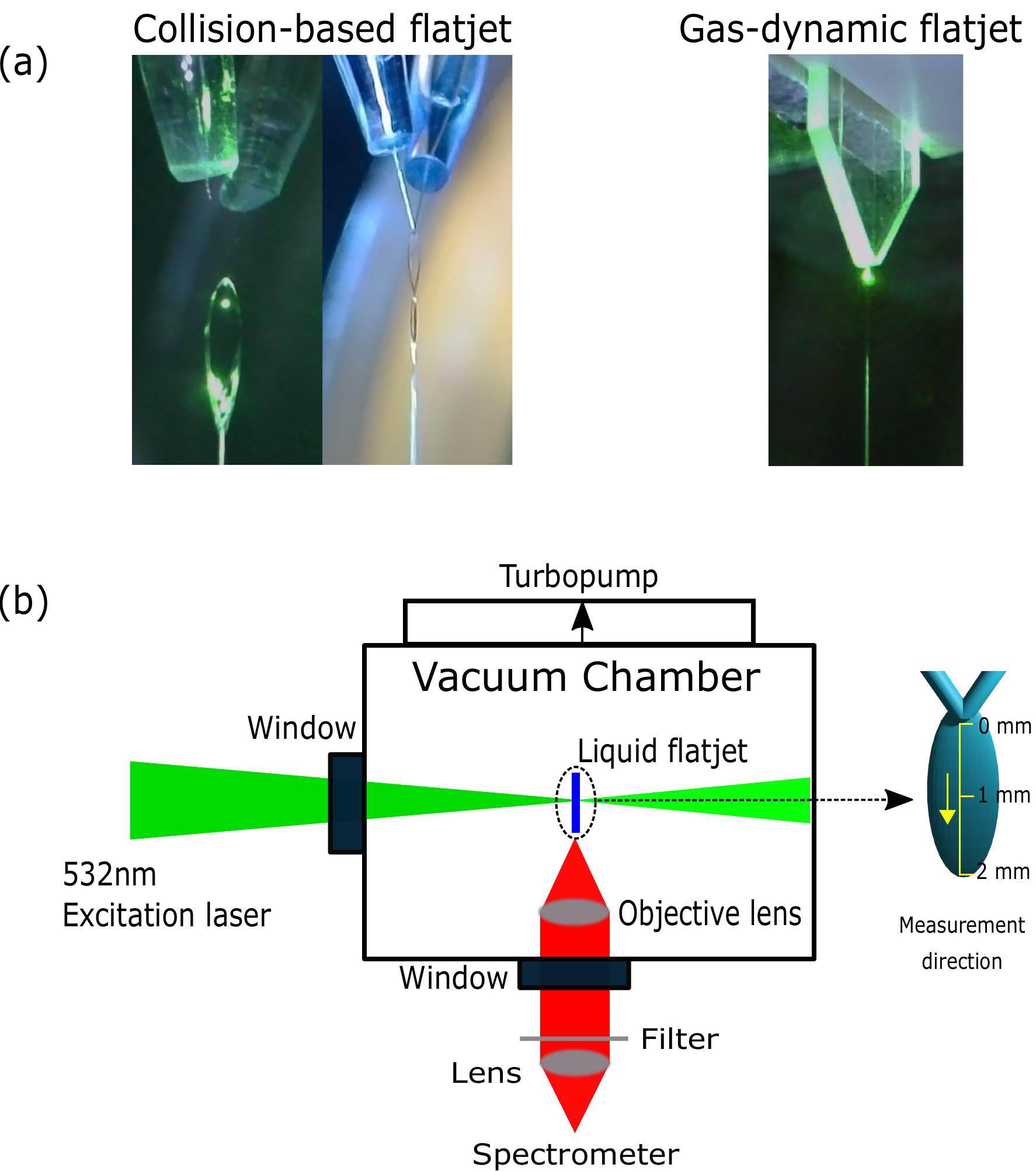} 
\caption{Raman thermometry of liquid flatjets in vacuum. Two different flatjet methodologies are used: collision-based flatjets formed by colliding two cylindrical jets together at an angle of 48$^{\circ}$, and gas-dynamic flatjets formed by using a focusing gas to compress a cylindrical liquid jet \SI{50}{\micro\metre} in diameter into a flatjet. (a) For the measurements, \SI{18}{\micro\metre} and \SI{60}{\micro\metre} orifices are used for collision-based flatjets, and \SI{50}{\micro\metre} orifice for gas-dynamic flatjets. (b) Top-down view of the sample chamber and a horizontal illustration of the flatjet. The flatjet is moved vertically within the chamber via an externally mounted manipulator and a cold trap attached below the sample chamber catches the remaining liquid.}
\label{fig:experiment_scheme}
\end{figure}

Fig. \ref{fig:experiment_scheme} illustrates our experimental scheme. As Raman excitation source, we used a Nd:YVO$_4$ cw laser (Coherent Verdi-V5) with a wavelength of 532\,nm and a maximum power of 2W. Heating of the flatjet by this excitation laser is negligible as measurements performed at different pump powers showed negligible variations in flatjet temperature. 

The Raman signal was collected at an angle of 90$^o$ with respect to the excitation beam by an objective lens within the liquid-jet chamber. The collimated signal was then sent out of the chamber via a window and focused by another lens onto the entrance slit of a flat field spectrograph (Princeton Instruments Acton SP2300). The resulting spectrum was recorded by an EMCCD camera (Andor iXon3).

For the calibration of the temperature measurement using Raman scattering, we used the flatjet under ambient conditions and simultaneously measured its temperature with a thermocouple. The sample, liquid water, was either cooled with a cold bath or heated with a warm bath and a heating device coupled to the nozzle holder, in order to get an absolute and accurate calibration curve. 

The Raman spectrum of liquid water around 3400 cm$^{-1}$ mainly consists in a superposition of five contributions (at 3050, 3200, 3400, 3500, and 3650 cm$^{-1}$), which correspond to the fundamental O-H stretching bands, vibrations and Fermi resonances.\cite{Monosmith1984,Suzuki2012, sun2013local} The intensity of each component changes relatively to the total intensity in response to temperature changes. For instance, with decreasing temperature, the integrated intensities of the 3200 and 3400 cm$^{-1}$ bands increase and that of the 3500 cm$^{-1}$ band decreases. 

To obtain the calibration curve, we normalized the recorded Raman spectrum at the midpoint 3300 cm$^{-1}$ and integrated the total band area above and below 3300 cm$^{-1}$. By plotting the ratio between these two integrated bands as a function of temperature, a calibration curve can be extracted by the expression\cite{wilson2004investigation,Smith2006,goy2018shrinking}:

\begin{equation}
    \frac{1}{T} =  C_{1}\ln\left(\frac{I_{<\Delta v}}{I_{>\Delta v}}\right) + C_{2}
\end{equation}

where $T$ is the temperature, $I_{<\Delta v}$ is the integrated intensity of the band below $\Delta v = 3300$ cm$^{-1}$ and $I_{>\Delta v}$ above it. $C_{1}$ and $C_{2}$ are empirical constants determined by calibration at known temperatures.

For the temperature measurements of ethanol, the anti-Stokes and Stokes signals could be recorded simultaneously so that the temperature was derived from the ratio of the two bands. A particularly well suited mode is the CCO symmetric stretch\cite{Emin2020} at 888.8 cm$^{-1}$. The temperature was thus extracted from the anti-Stokes/Stokes ratio:

\begin{equation}
    \frac{I_{AS}}{I_{S}} =  \left(\frac{V_{l} + V_{\Delta v}}{V_{l} - V_{\Delta v}}\right)^{4} \exp\left(\frac{-hV_{\Delta v}}{kT}\right)
\end{equation}

where $T$ is the temperature, $k$ the Boltzmann’s constant, $h$ the Planck’s constant, $V_{l}$ the frequency of the laser and $V_{\Delta v}$ the frequency of the CCO symmetric stretch mode (center frequency).

\clearpage\section{Results and Discussion}

\begin{figure}[htbp]  
\centering
\includegraphics[width=12cm]{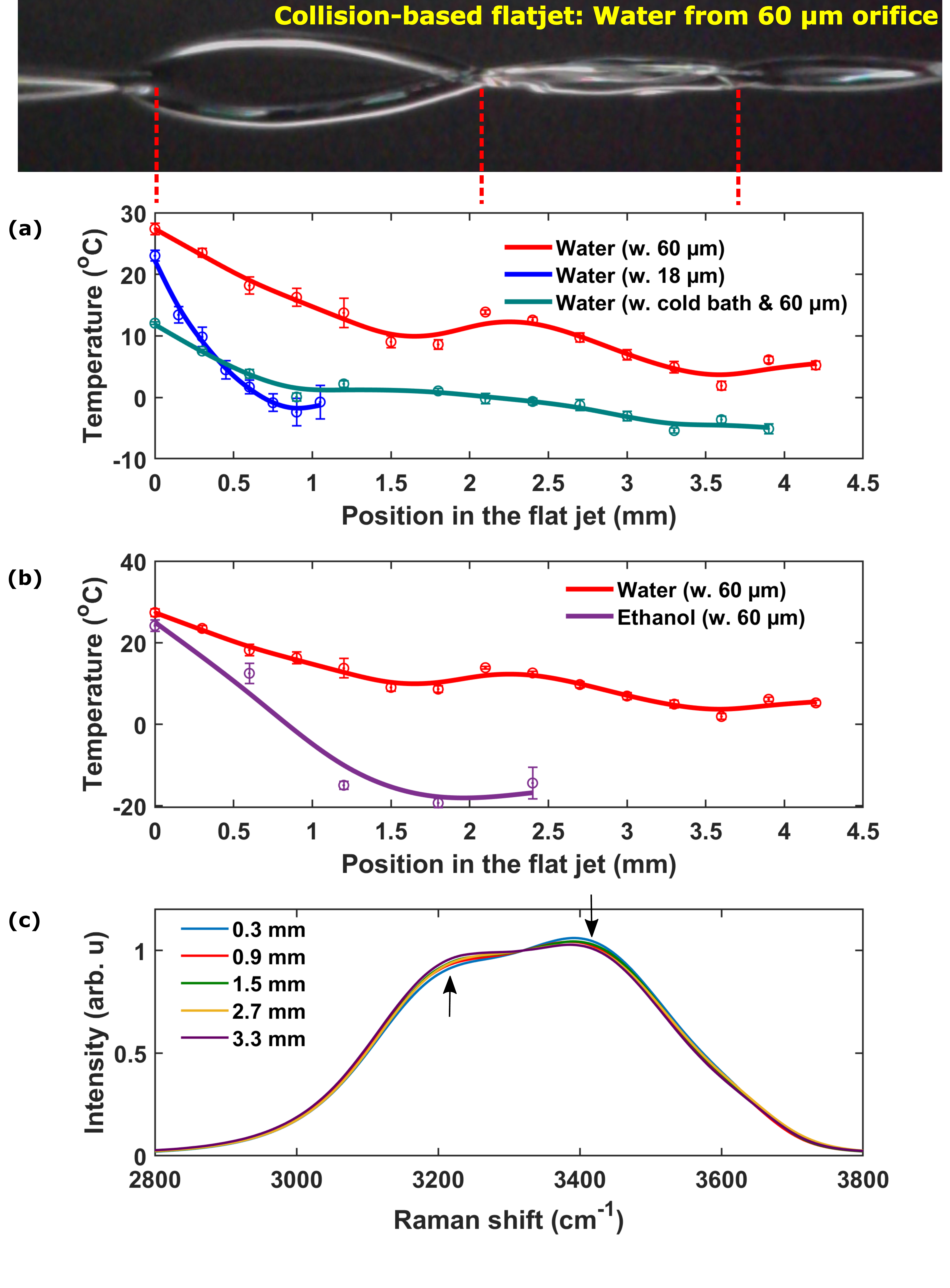}
\caption{Measurements with collision-based flatjets of different samples under different initial temperature, flowrate and orifice size, starting from an initial position defined as 0mm just below the collision point as indicated by the red dashed line. a) Thermal profiles of water using \SI{18}{\micro\metre} nozzles at 2 ml/min (Blue), water using \SI{60}{\micro\metre} nozzles at 5 ml/min (Red), and water using \SI{60}{\micro\metre} nozzles at 5 ml/min with an initial temperature of ~12 $^{\circ}$C maintained by placing the sample in an external cold-bath (Green). b) Thermal profile of ethanol using \SI{60}{\micro\metre} nozzles at 5 ml/min (Purple) as measured via the stokes/anti-stokes ratio. c) Raman spectra of liquid water at different positions along the flatjet of water in cold-bath.}
\label{fig:data}
\end{figure}

Fig. \ref{fig:data} shows the temperature profile along the collision-based flatjets. The origin is set at the collision position. For liquid water flowing from the \SI{18}{\micro\metre} nozzles (blue curve), the maximum cooling of 25$^{\circ}$C is obtained just before the end of the flat sheet, leading to a final temperature of $\sim$-2$^{\circ}$C. Due to the small nozzle diameter, only one flat sheet of about 1 mm length is formed.

For larger nozzle diameters (\SI{60}{\micro\metre}), the thickness of the flat sheet increases and the flow is more stable. This allows the formation of a second subsequent flat sheet, as shown in Fig. \ref{fig:data}a (red and green curves). The first sheet is about 2 mm long and has a minimum temperature of $\sim$8 $^{\circ}$C. As it collapses into a single jet, temperature increases because of water masses from the rim of the flat sheet that now mixes with the central region. On the rim, the flatjet is thicker, less prone to evaporative cooling and warmer than the center. The second flat sheet is slightly shorter (about 1.5 mm), with a minimum temperature of $\sim$2$^{\circ}$C. The overall cooling rate is thus lower for the \SI{60}{\micro\metre} nozzles (10 K/mm) than for the smaller \SI{18}{\micro\metre} nozzles (28 K/mm). This is consistent with the results from Ekimova et al.\cite{Ekimova2015}, who estimated that evaporation yielded a mass loss of $5\%$ under conditions comparable to ours, so that the thickness of the second flat sheet is still thicker than the single flat sheet from the \SI{18}{\micro\metre} nozzles. It has to be noticed, however, that the flowrates had to be adapted for obtaining stable jets for the \SI{60}{\micro\metre} (5 ml/min) and for the \SI{18}{\micro\metre} (2 ml/min) nozzles.

Interestingly supercooled water (-2 $^{\circ}$C) can thus be produced, when using collision-based flatjets with small nozzles. Supercooled water can, however, also be produced using \SI{60}{\micro\metre} nozzles, if the water is previously cooled at around 12 $^{\circ}$C instead of room temperature, as shown by the green plot in Fig. \ref{fig:data}a. Moreover the length of the cold region zone is extended over several mm.

As compared to water, ethanol experienced a much stronger cooling rate (Fig. \ref{fig:data}b), due to its faster evaporation. Even with the \SI{60}{\micro\metre} nozzles, an average cooling rate of $\sim$ 29 K/mm is observed, which is 3 times faster than water under the same conditions. Only one sheet can be formed due to the loss of matter, and the lowest temperature found in this sheet reaches -20 $^{\circ}$C. For the same reason, a stable flatjet of alcohol was difficult to produce when using the smaller nozzles.

\newpage
\begin{figure}[htbp] 
\centering
\includegraphics[width=14cm]{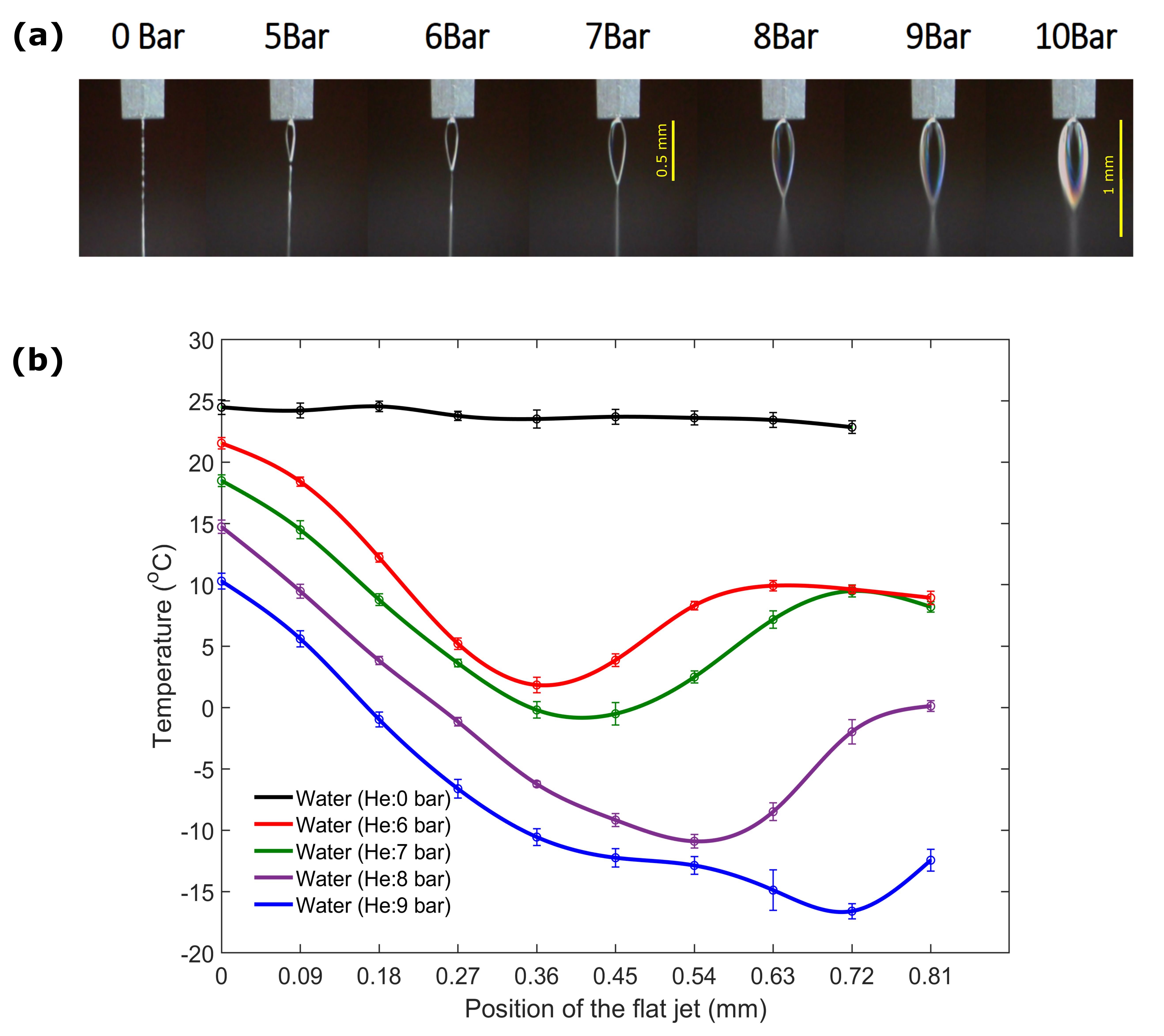}
\caption{Measurements with a microfluidic gas-dynamic chip nozzle. a) Gas-dynamic flatjets generated with different focusing gas pressure from 5-10 bar of helium, at a constant flowrate of 1 ml/min. A lower liquid flowrate is required for operation compared to collision-based flatjets, therefore much smaller flatjets are formed.\cite{Koralek2018} b) Thermal profiles of the gas-dynamic flatjets, at an initial position defined as 0 mm just below the nozzle tip.}
\label{fig:data_gasjet}
\end{figure}

\begin{table}[ht]
  \caption{Gas-dynamic flatjets: experimental values of the exit temperature (temperature at the highest point of the flatjet) and average cooling rate for different focusing gas pressure. The maximum jet velocity of the flatjet is 8.5 m/s.}
  \includegraphics[width=\linewidth]{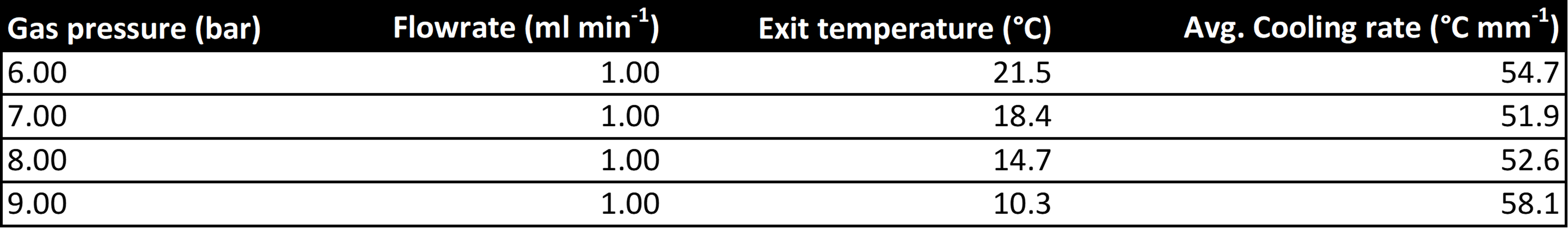}
  \label{tbl:exp_values_gasjet}
\end{table}

For the gas-dynamic flatjet, as shown in Fig. \ref{fig:data_gasjet}, the measurements start at approximately 0.1 mm just below the nozzle. Note that the focusing gas starts compressing the liquid sample in the nozzle tip itself. By holding the flowrate constant at 1 ml/min and increasing the gas pressure, one observes an increase in the flatjet size up to 10 bar, at which point the flatjet starts disintegrating. The overall flat sheet is also much shorter than for the colliding-jet source, with a maximum length of about 1 mm. As shown in Table \ref{tbl:exp_values_gasjet}, for gas pressures of 6-9 bar, the cooling rates reach as much as 54.7, 51.9, 52.6 and 58.1 K/mm, respectively and an initial exit temperature of 21.5, 18.4, 14.7 and 10.3 $^{\circ}$C, respectively. While the average cooling rate appears to have little-to-no dependence on the focusing gas pressure, the initial exit temperature is strongly dependent on it. This suggests that while the use of a focusing gas limits the effect of evaporative cooling, heat is still lost through conductive cooling between the gas and liquid. 

\section{Model Calculations}

\begin{figure}[htbp]
\centering
\includegraphics[width=14cm]{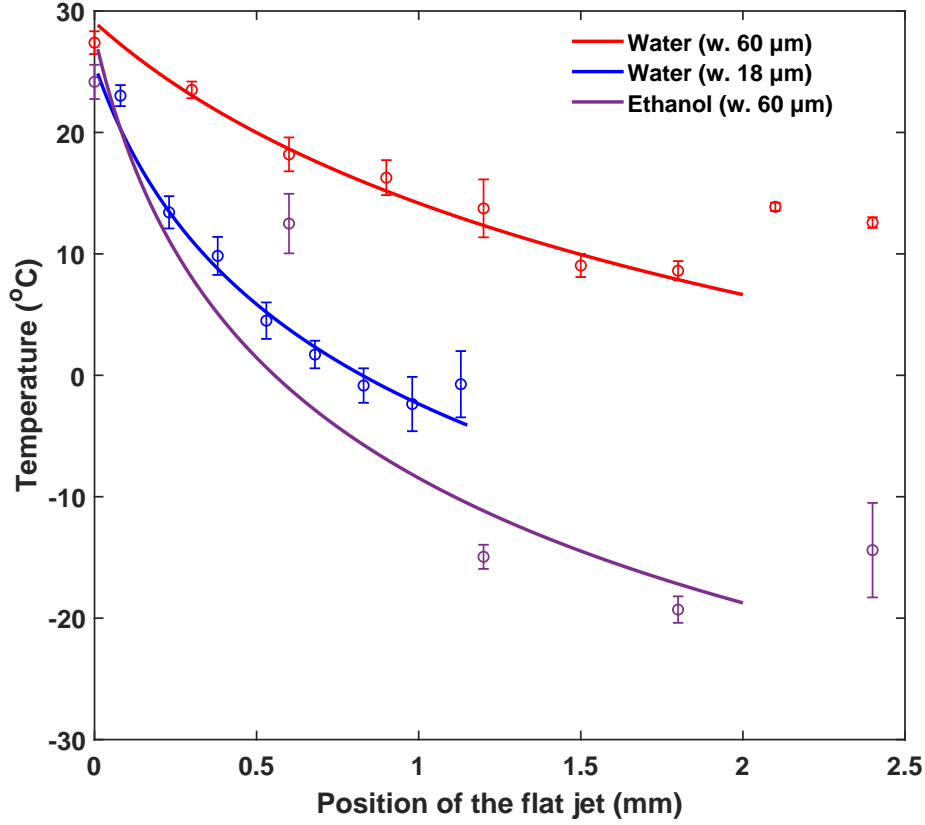}
\caption{Fitting of experimental data through numerical integration of Eq. \ref{cooling rate_model}. Blue: water with \SI{18}{\micro\metre} nozzles. Red: water with \SI{60}{\micro\metre} nozzles. Purple: ethanol with \SI{60}{\micro\metre} nozzles.}
\label{fig:theory_fit}
\end{figure}

To explain the thermal evolution of the colliding flatjet, we developed a basic model to fit our measurements. No modeling was performed on the gas-dynamic flatjet experiment, due to its complex geometry and some unknown geometrical parameters in this commercial device. To begin, we use the Hertz-Knudsen equation to describe the mass flux $J_{HK}$ from a liquid surface, which is generally given in the form \cite{knudsen1935,Persad_Ward_2016}:

\begin{equation}\label{HK eq}
    J_{HK} =  \frac{1}{A}\frac{dN}{dt}= \frac{\alpha_{s}}{\sqrt{2\pi m k_{B}}}\left(\frac{\sigma_{e} P_{S}}{\sqrt{T_{I,L}}} - \frac{\sigma_{c} P_{V}}{\sqrt{T_{I,V}}}\right)
\end{equation}

where $A$ is the area, $N$ is the number of molecules, $k_{B}$ is the Boltzmann constant, $T_{I,L}/T_{I,V}$ is the interfacial liquid/vapor temperature, $m$ is the mass of a molecule, $P_{S}$ is the saturation vapor pressure, $P_{V}$ is the gas pressure, $\sigma_{e}$ is the evaporation coefficient, $\sigma_{c}$ is the condensation coefficient, and $\alpha_{s}$ is the sticking coefficient of the gas onto a surface. 

In addition, the change in temperature $dT$ due to the phase transition (evaporative cooling) of $dN$ molecules is given by

\begin{equation}\label{enthalpy change}
   dT = \frac{H_{v}}{N_{A} C_{m} M V} dN
\end{equation}

where $H_{v}$ is the evaporation enthalpy, $N_{A}$ is the Avogadro constant, $C_{m}$ is the molar specific heat, $M$ is the molar density, $V$ is the volume of the liquid.

Combining Eq. \ref{HK eq} and Eq. \ref{enthalpy change} yields the following expression for the change in temperature per unit length of the liquid jet in the vertical z-axis (i.e. cooling rate):

\begin{equation}\label{cooling rate}
\begin{split}
   \frac{dT}{dz} &= -\frac{H_{v}}{N_{A} C_{m} M V} \frac{dN}{dt}\frac{dt}{dz} \\
   &= -\frac{H_{v}}{N_{A} C_{m} M V} \frac{2A J_{HK}}{v_{z}} \\
   &= -\frac{2H_{v}}{N_{A} C_{m} M l_{j}(z)} \frac{J_{HK}}{v_{z}}
\end{split}  
\end{equation}

where $l_{j}$ is the thickness of the jet and $v_{z}$ is the velocity of the jet along the z-axis.

The thickness of the jet has been estimated by Hasson and Peck\cite{Taylor1960,HassonPeck_Thickness} as:

\begin{equation}\label{thickness}
l_{j}(z) = \Lambda \frac{d_{o}^2}{4z} \frac{\sin^3(\phi)}{(1+\cos(\theta) \cos(\phi))^2} = \frac{l_{0}}{z};
\end{equation}

where $z$ is the distance from the collision point, $d_{o}$ is the orifice diameter, $2\phi = 48^{\circ}$ is the collision angle, $\theta=0^{\circ}$ is the azimuthal angle, $\Lambda$ is an unitless proportionality constant, and $l_{0}$ is the overall proportionality constant ($m^2$) at fixed $\phi$ and $\theta$. From white-light interferometric measurements of the flat jet thickness, a value of $\Lambda = 124$ is obtained as the best fit. Equation \ref{thickness} predicts a linear decrease of the jet thickness with the distance from the collision point. However, it has been observed in our and other experiments\cite{Ekimova2015,Luu2018,yin2020} that not only is the boundary rim of the flat sheet substantially thicker than its center, the size of the flat sheet also varies with the flowrate of the liquid jet, which is not captured by Eq. \ref{thickness}. While more complex models that take into account these effects exist\cite{bush_hasha_2004,ChooKang2007}, we decided that for our measurements along the center of the sheet, Eq. \ref{thickness} was reasonably applicable once minor corrections were made.

Although the Hertz-Knudsen (HK) model is perfectible\cite{Persad_Ward_2016, Holyst2015}, we used a simplified evaporation flux model $J_{\gamma}$, based on the HK equation, where the equilibrium/saturation vapor pressure $P_{v}$ is given by the Clausius–Clapeyron equation:

\begin{equation}\label{evaporation_model}
\begin{split}
J_{\gamma} &= \frac{\gamma P_{v}}{\sqrt{2\pi m k_{B} T}} \\ &=  \frac{\gamma P_{ref}}{\sqrt{2\pi m k_{B} T}} \exp\left[-\frac{H_{v}}{k_{B} T_{ref}} (\frac{T_{ref}-T}{T})\right] \\ &= \gamma \frac{J_{0}}{\sqrt{T}} \exp\left[-\frac{H_{v}}{k_{B} T_{ref}} (\frac{T_{ref}-T}{T})\right]
\end{split}
\end{equation}

where $T_{ref} = 298.15$K is the reference temperature, $P_{ref}$ is the vapor pressure at $T_{ref}$, $\gamma$ is an unitless proportionality constant and $J_{0} = P_{ref}/\sqrt{2\pi m k_{B}}$. From Eq. \ref{evaporation_model}, we define our final simplified fitting model by: 

\begin{subequations}
\begin{equation}\label{cooling rate_model}
\begin{split} 
   \frac{dT}{dz} &= -\frac{2H_{v}}{N_{A} C_{m} M l_{j}(z)} \frac{J_{\gamma}}{v_{z}} F(z) \\ &= -2\gamma \epsilon \frac{H_{v}}{N_{A} C_{m} M l_{0} v_{z}} \frac{J_{0}}{\sqrt{T}} \exp\left[-\frac{H_{v}}{k_{B} T_{ref}} (\frac{T_{ref}-T}{T})\right] \\ &= -\frac{\alpha C}{\sqrt{T}} \exp\left[-\frac{H_{v}}{k_{B} T_{ref}} (\frac{T_{ref}-T}{T})\right]
\end{split}   
\end{equation}
with 
\begin{equation}
F(z) = \frac{\epsilon}{z} 
\end{equation}
\begin{equation}\label{cooling coefficient}
   C_{} = \frac{H_{v} J_{0}}{N_{A} C_{m} M l_{0} v_{z}} 
\end{equation}
\end{subequations}

where $\alpha = 2\gamma \epsilon$ is an overall proportionality constant with dimension $m$, with $\epsilon$ a correction constant, and $C$ is the cooling coefficient. $F(z)$ is a correction function for the flatjet's geometric properties. The fits to the experimental data obtained from numerical integration of Eq. \ref{cooling rate_model} are presented in Fig. \ref{fig:theory_fit}.

\begin{table}[H]
  \caption{Proportionality constants $\alpha$ are unknown variables obtained from fittings of the experimental data with numerical integration of Eq. \ref{cooling rate_model}. Cooling coefficients $C$ are known variables that can be calculated from experimental parameters and provide an estimation for the cooling rate of an unknown sample relative to a known one.}
  \includegraphics[width=\linewidth]{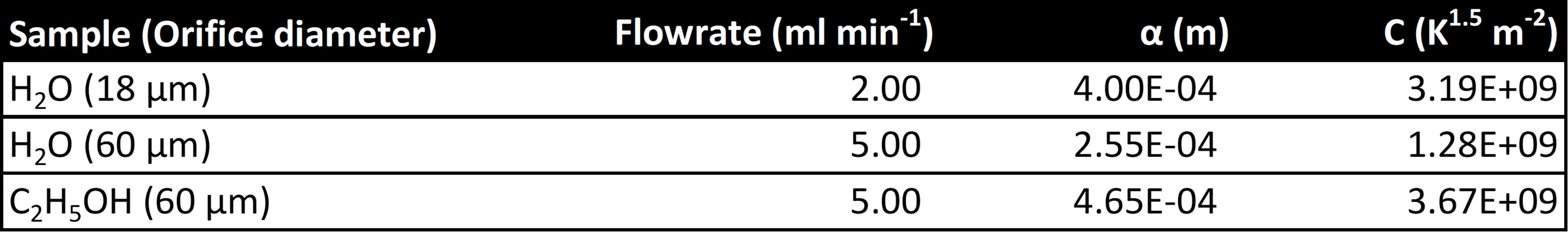}
  \label{tbl:coefficients}
\end{table}

As listed in Table \ref{tbl:coefficients}, different values of $\alpha$ and cooling coefficient $C$ are given for the three different cases of water from \SI{18}{\micro\metre}, water from \SI{60}{\micro\metre} and ethanol from \SI{60}{\micro\metre} orifices. The ratio of the average experimental cooling-rate for different measurement sets can be approximately expressed as a ratio of cooling coefficients $C'/C$.

According to Eq. \ref{cooling coefficient}, although the cooling coefficient is not directly dependent on the nozzle orifice-diameter (given Eq. \ref{thickness}), it is inversely proportional to the jet velocity and therefore the flowrate, which is orifice-diameter dependent. Experimentally, flowrates of 2 and 5 ml min$^{-1}$ were used for \SI{18}{\micro\metre} and \SI{60}{\micro\metre} orifices respectively, with an experimental cooling-rate ratio of $\sim$3, compared to $C^{H_{2}O}_{18\mu m}/C^{H_{2}O}_{60\mu m} = 2.5$. Given that the flatjets are only stable within a small flowrate range for a given orifice size, and that their areas depend on the flowrate, it was experimentally difficult to observe a flowrate dependence for a fixed orifice. The cooling coefficient ratio for ethanol/water is $C^{EtOH}_{60\mu m}/C^{H_{2}O}_{60\mu m} = 2.87$, which is similar to the experimental cooling-rate ratio of $\sim$$2.9$. 

For the thermal evolution of the gas-dynamic flatjet, an alternative model to the colliding jet would need to be developed. Such a model would need to take into account not just evaporative cooling, but also the strong likelihood of conductive cooling between the compression gas and sample liquid. This may be explored in a future study. From the measurements, it can be concluded that the focusing gas does not prevent the liquid sample from cooling. The use of the gas-compressed flatjet to reach the deeply supercooled regime of water is left for future studies.\cite{Grisenti2018}

\section{Conclusion}
In conclusion, we have measured the thermal profiles of two different types of liquid flatjet systems in vacuum utilizing optical Raman spectroscopy and observed the temperature dependence on the material vapour pressure, orifice size, flowrate and initial temperature of the liquid sample. Lower temperatures are accessible in water with the gas-compressed flatjet than the collision-based flatjet. Based on the results, we developed a simplified, empirical model to describe the effect of evaporative cooling on collision-based flatjets. In addition, we demonstrated that liquid water in the flatjets can enter the supercooled regime, which opens up future investigations of the supercooled water using transient XAS, HHG spectroscopy, electron diffraction, attosecond spectroscopy\cite{Jordan2020} and many more. More generally, the measurements reported herein will facilitate the first systematic temperature-dependent studies of liquid-phase systems using these novel techniques.  

\section*{Acknowledgements}
Acknowledgements: It is our pleasure to thank Andreas Schneider, Mario Seiler, Andres Laso, and Markus Kerellaj for their contributions to the construction of the experiment and  Michel Moret for his precious technical assistance.
This work was supported by ETH Zurich, the Swiss National Science Foundation (SNSF) through the NCCR-MUST and project (No 20021-172946) and an ERC Consolidator Grant (No 772797-ATTOLIQ). T.B. acknowledges financial support from a Marie-Curie fellowship grant agreement No 798176. Z.Y. acknowledges financial support from an ETH Career Seed Grant No SEED-12 19-1/1-004952-000.

\section*{Author Declarations}
\subsection*{Conflict of Interest}
The authors have no conflicts to disclose.

\section*{Data Availability}
The data that support the findings of this study are available from the corresponding
authors upon reasonable request.

\bibliography{Draft_Raman_V5.bib}

\end{document}